# Languages for Mobile Agents

Steven Versteeg

Supervisor: Leon Sterling

## Abstract

**Mobile agents represent a new model for network computing. Many different languages have been used to implement mobile agents. The characteristics that make a language useful for writing mobile agents are: (1) their support of agent migration, (2) their support for agent-to-agent communication, (3) how they allow agents to interact with local resources, (4) security mechanisms, (5) execution efficiency, (6) language implementation across multiple platforms, and (7) the language's ease of programming of the tasks mobile agents perform.**

# 1. Introduction

Mobile agents are an emerging technology that promise many benefits in network computing.  A mobile agent is a program that can migrate from one computer to another for remote execution.  Many different languages have been used to implement mobile agents.  This thesis examines the characteristics required for a language to be useful for writing mobile agents.  Telescript, Java, Agent Tcl and Obliq are examples of mobile agent languages that are examined to determine what makes them useful.

# 2. Background

Mobile agents are in the process of graduating from being limited to research systems to being a practical technology in network computing.  Mobile agents are computer programs which may migrate from one computer to another on a network.  On migration, the agent suspends at an arbitrary point before migrating, and restarts execution at that point when it resumes execution on the target computer. [Ven97] [Gra95a]  The word *agent* is used to describe a very broad range of programs.  The exact definition of the word is vague.  Often associated with agent is the implication that the programs are persistent, autonomous and interact with their environment.  Others define agent to simply mean a program that does a task on behalf of a user.  Both these sets of properties are generally true of mobile agents.  In the context of this discussion mobile agent is simply a program that can migrate from one computer to another.  Any other conflicting definitions of the word agent should be ignored.

The main advantage of mobile agents is that they can bring a program closer to the information resources.  The mobile agent paradigm stipulates that the server should provide set of basic services.  The client uses the services provided by the server by dispatching a program, that is a mobile agent, to the server.  The mobile agent makes use of the server's basic services, in the way that its owner intends.  Mobile agents provide no new functionality that cannot be achieved with traditional client-server interaction, such as remote procedure call (RPC).  However, they make implementing any new functionality much easier.  The fundamental advantage is they provide a layer of abstraction, between the services provided by the server and the way they are used.  For a further introduction into mobile agents, and a critical analysis of their advantages, the reader is referred to Harrison, Chess and Kershenbaum, Mobile agents: Are they a good idea? [HCK95]

In the context of a discussion of what languages are useful for writing mobile agents, it is necessary to know what type of applications are being written.  While mobile agents are not new, they are still in the process of moving from research systems to mainstream computing.  Mobile agents are expected to be able to roam over heterogeneous networks, such as the Internet.  The types of applications that mobile agents are envisioned to be used for are:

- Search and gathering applications.  Mobile agents roam across the network, searching the servers' resources for a specific piece of information.
- Monitoring programs.  A mobile agent sits on a server monitoring information, until a condition is met.
- Electronic commerce.  Mobile agents act as representatives of a user, and search for and buy products on the user's behalf.
- Distributed computing.  Mobile agents can be used as mechanisms to distribute computation across the network.

This simple example illustrates how mobile agents can be usefully applied.  The problem: the user needs to be informed, exactly when the stock price of BHP rises above a certain threshold.  The mobile agent solution:  A mobile agent is dispatched from the user's computer to a stock exchange server, that provides a feed of the course of sales in real time.  The agent sits at the server and monitors the sales.  When it finds a sale with a price above the threshold, it migrates back to the client computer and informs the user.  The whole scenario may take days or even weeks to complete.  Only two network communications were made.  One to send the agent to the stock exchange server, and one to send it back again. Consider the alternative ways of implementing this functionality.  One way is to send all the course of sales information from the stock exchange server to the users computer.  At the user's computer, a local program monitors the sale price.  This solution involves thousands of network communications.  Another solution is to use Remote Procedure Call (RPC).  A program runs on the user's computer that polls BHP's price at certain time intervals through a RPC.  This alternative is causes less network traffic, but still much more than for the mobile agent solution.

A mobile agent is merely a program.  The mobile agent requires an environment on potential hosts to run on.  All agent systems have an *agent server* running on all potential host machines.  The agent server acts like an operating system for mobile agents.  The agent server is responsible for: (1) providing an environment for the agent to run in; (2) transferring and receiving agents to and from different agent servers; and (3) implementing an API for messaging between agents and agent transfer requests.  It is also the responsibility of the agent server to protect the host computer from hostile mobile agents.

Mobile agents programs are only able to run on hosts that have an execution environment that interprets the language they were written in.  There generally needs to be a separate kind of execution environment for each language.  It is possible for an agent server to be able to support more than one language, however there are presently many competing and incompatible types of agent servers, each only capable of interpreting at most a few languages.  Some agent operating systems (or types of agent servers) are Ara [RP97], Tacoma [JRS94], and the Knowbot Operating System [Hyl96].  Agent Tcl and Telescript each have their own agent operating systems.  The many different Java-based agent systems also each require special agent server.

This thesis is concerned with programming languages for writing mobile agents rather than the operating systems they execute under. The implementation of agent servers is only discussed if it directly affects the programs that can be implemented.

# 3. Languages Used to Write Mobile Agents

In theory any language can be used to implement mobile agents.  The only necessary requirement is that the language is supported by an execution environment on the host.  A wide variety of languages have been used to write mobile agents, some in research systems, some in prototype commercial systems.  Some languages such as Obliq and Telescript have been specifically designed for writing mobile agents.  There are also many mobile agents being written in general purpose languages extended with a special library.  Below is a brief description of some of the languages that have been used to write mobile agents.

**Telescript** - A proprietary system developed by General Magic. [Whi96]  The Telescript language has been specifically designed for implementing mobile agent systems.  Telescript was designed with the vision for the computer network become a programmable platform.  General Magic's ambition was for Telescript to become for communications what Postscript is for printing.  Contrary to the name, Telescript is not a scripting language.  It is a complete object oriented language.  Telescript supports objects, classes and inheritance.  The object oriented model and the syntax is in many way similar to that of C++.  Telescript has a library of built-in classes for writing mobile agents.  There are special classes for *agents* and *locations*.  Agents are a base class for mobile agents.  Locations are objects that represent sites.  The Telescript language has a set of built-in commands for agent migration and inter agent communication.  The Telescript system includes notions of which authority the agent is representing.   Telescript programs are compiled into a portable intermediate representation, called *low Telescript*, analogous to Java byte code.  Telescript programs can run on any computer with a Telescript execution engine.  The Telescript execution engine was designed to be able to run on even small communication devices.  The Telescript language has had a great influence on the development of mobile agents, and mobile agent languages.  It was General Magic who first coined the term *mobile agent*.

**Java** - Java is a general purpose language.  Despite its relatively young age, it is already establishing itself as the de facto standard for developing internet and intranet applications.  Java is an object oriented language.  It uses the classes  object oriented model.  Its syntax is similar to that of C and C++. While Java was not specifically designed for writing mobile agents, it has most of the necessary capabilities for mobile agent programming.  Java is multi-threaded.  Java programs are compiled to Java byte codes, binary instructions for the Java Virtual Machine.  Java programs are able to run on any platform with a Java Virtual Machine interpreter.  This makes Java programs highly portable.  The Java libraries have good support for communication procedures.  Java has been used as the basis for many implementations of mobile agent systems.  Nearly all of the systems make use of Java 1.1's RMI (Remote Method Invocation).   Some systems of note include:

- IBM's **Aglets** - under development by IBM Research Centre, Japan.  An *aglet* is a mobile agent.  All aglets are derived from an abstract class called Aglet.  Aglets uses an event driven approach to mobile agents, that is analogous to the Java library Applet class. [KZ97]  Each aglet implements a set of event handler methods that define the aglets behaviour.  Some of these methods are:
    - OnCreation() -- called when a new aglet is created.
    - OnDispatch() -- called when an aglet receives a request to migrate.
    - OnReverting() -- called when the aglet receives a request from its owner to come home.
    - OnArrival() -- called after an aglet is dispatched
- General Magic's **Odyssey** - A mobile agent system under development by General Magic, that attempts to achieve the functionality of Telescript, using Java.
- ObjectSpace's **Voyager** - The Voyager system's model of mobile computing is very similar to that of Obliq.  The system provides a mechanism for converting objects into a distributed objects.  This allows objects at remote sites to be semantically treated in the same way as objects at the local site.  Objects can be easily copied between remote sites. [KZ97]

**Obliq** - Obliq is an experimental language under development by Digital Equipment Corporation's Systems Research Center.  Obliq is a lexically scoped, object-based, interpreted language that supports distributed computation. The language supports objects, but not classes.  It uses the prototype-based model [Bor86] of object-oriented programming.  New objects can be created directly, or cloned from other objects. Obliq uses runtime type checking. Obliq has built-in procedures for importing and exporting procedures and objects between machines. Obliq adheres to lexical scoping in a distributed context. When procedures and objects are dispatched to a remote site for execution, any references they contain point to the same objects as on the machine from which they were dispatched. [Car95] [BC96]

The Obliq distributed semantics is based on the notions of *sites*, *locations*, *values* and *threads*. A site is a computer on the network. A location is a memory address on a site that stores a value. A value can be of a basic type or an object. Threads are virtual sequential instruction processors. Threads may be executed concurrently on the same site or at different sites. Values may be transmitted over the network. When an object is transmitted, basic values are copied exactly. Locations that the object contains are copied, such that they point to the same address on the same site, at the destination site as they did at the original site.

Obliq's semantics of network computing is fundamentally different to the other languages considered. Where as other languages see each computer as independent worlds that can communicate with each other through the network, Obliq treats the network as a single computer with sites as components.

**Agent Tcl** - Agent Tcl [Gra95b] is a mobile agent system being developed by Dartmouth College. The Agent Tcl language is an extension of the Tool Command Language (Tcl), the language originally developed by Dr. John Ousterhout. The Agent Tcl extensions add commands for agent migration and message passing. The extra commands give Agent Tcl scripts similar mobility capabilities to Telescript. Agent Tcl uses a modified Safe Tcl [OLW96] interpreter to execute scripts.

**Perl 5** - Penguin is a Perl 5 module with functions enabling the sending of Perl scripts to a remote machine for execution and for receiving perl scripts from remote machines for execution. The scripts are digitally signed to allow authentication and are executed in a secure environment. Mobile agents written in Perl are restricted in that they must always restart execution at the same point. There is also no support for agents saving their state on migration. A new Agent Module v3.0 is being created to give Perl 5 more sophisticated mobile agent capabilities. The extra features include giving agents the ability to save their state on migration.

**Python** - Python is an object-oriented scripting language. The Corporation for National Research Institution, uses Python as a language for implementing Knowbot programs. [Hyl96]

This is by no means a complete list of the languages being used for mobile agents. For a more complete list, the reader is referred to Kiniry and Zimmerman [KZ97].

The languages that will be mainly considered in the following discussions are Telescript, Java, Agent Tcl and Obliq. Collectively, these languages represent most of the approaches presently taken to languages for mobile agents.  Aglets will be most referred to of the Java libraries.  The reason for this is the techniques associated with the other two Java libraries mentioned are represented by Telescript and Obliq.

# 4. Characteristics of Languages for Mobile Agents

Any language used to write a mobile agent must be able to support the following:

- agent migration,
- communication between agents,
- access to server resources,
- security mechanisms,
- appropriate efficiency
- the ability to run on multiple platforms
- ease of programming for writing mobile agent application.

How well the language is able to support these stipulates the usefulness of the language for writing mobile agent applications.

## 4.1 Migration

The agent language must be able to support an agent migrating.  Ideally, it should be possible to suspend an agent's execution at any point, save the state, including the heap, the stack and even the registers, move the agent to another computer, and restart execution, with the agents execution state exactly restored.

Telescript has built-in support for agent migration.  Agents may move to any location with the go statement.   Upon the execution of this command, the agent is transported to the target site, where it continues execution from the line after the go statement.  All the agents properties and the program execution state, including those of local variables in methods and the program counter, are restored exactly.  The agent migration is process is handled completely by the Telescript operating system.  The programmer does not need to worry about saving the relevant state information just before migration. [Whi96]

Agent Tcl uses a similar migration model to that of Telescript. The built-in statement for agent migration is called agent_jump. As with the Telescript go, when this statement is issued the execution environment handles the transportation of the agent, and restores the agent execution state. Since the Tcl language provides absolutely no support for capturing program state, this is an Agent Tcl extension of the language.

Java was not specifically designed for implementing mobile agents so it does not have in built-in support for migration. Saving the program state in Java is much more difficult. Java's security architecture makes it impossible to directly save the virtual machine execution state. However Java 1.1 supports class serialization. Serialization allows an entire class instance to be written to file, including the object's methods, attributes and their values. Serialization will not save the program stack, that is, the values of local variables in methods. The Java virtual machine does not allow the explicit referencing of the stack, for security reasons. Workarounds have been developed for saving the program stack state. In Aglets, each aglet implements a method called onDispatch(). This method is called when an aglet receives a request to migrate. The request may have come from the aglet itself or from another process. In this method, the programmer must define a procedure for placing everything an aglet needs to restore its state on the heap. The aglet is then serialized and transported to its destination. [Ven97a]

There are advantages to Telescript and Agent Tcl's built-in support for agent migration. In Telescript it is possible to migrate from any point in the program, including in the middle of method calls. In Java the agent program must be structured so that everything needed to restore execution state is stored in the heap, before migration. It is left to the programmer to make sure that all variables are correctly saved. In Telescript and Agent Tcl, the implementation of agent migration is completely hidden from programmer. This is a source of error that Telescript programmers do not need to worry about.

Obliq takes a different view of agent migration. In Obliq, an agent can be written as a procedure that takes a state object as an argument. A site can make its execution engine available for threads at other sites to use. A procedure can be executed at a remote site, by passing the name of the procedure as a parameter to the execution engine. The following code fragment shows how an agent can be sent to another site for execution. [Car95]

let state = { ... };                    *(define agent state)*

let agent = proc(state, arg) ... end;   *(define agent procedure)*

*(get a handle to remote site execution engine)*

```
let remoteSite = net_import("RemoteServer", Namer);
```

*(Execute the agent at the remote site.)*

```
remoteSite(proc (arg) agent(copy(state), arg) end)
```

## 4.2 Agent communication

The agent language must allow agents to communicate with each other.

In Telescript agents communicate by holding meetings. An agent can request a meeting with another agent at the same place, that is the same execution environment. The Telescript system passes the meeting request to the relevant agent. Every Telescript agent must implement the operation meeting. This is called when an agent receives an invitation to hold a meeting. The implementation of the meeting method contains the agents negotiating strategies, which may include rejecting holding a meeting under certain conditions or with certain types of agents. [Whi96]

Agent Tcl provides extensions to the Tcl language for agent communication. These extensions allow agents to communicate through either asynchronous message passing, or through remote procedure calls. [Kot97]

Java has no built in support for agent communication. In Aglets, each Java agent has a proxy object. Communication from one agent to another happens through the proxy. This is to protect the agent objects from being directly modified. The proxy object provides a set of methods for communicating to the represented object. These include requests for aglets to take actions, such as migration, cloning, destroying and suspending. There are also two methods for sending synchronous and asynchronous messages to the aglets. [Ven97a]

## 4.3 Interface to server resources

The fundamental purpose of mobile agents is to get the program closer the source of the information. The agent implementation language must provide an easy way to access the resources on the host machine.

In Telescript, local resources are treated as another agent. There is an agent present at the server to represent the local resources. This model provides an elegant and consistent interface to local resources at different computers, but it requires writing a Telescript wrapper. [Whi96]

Obliq has categories different types of services provided by a site. A program may request a list of the services provided by a site in a particular category.

Agent Tcl and Aglets use a similar method to interacting with local resources to Telescript. In Aglets, an aglet is associated with an AgletContext object. This object describes the environment that the aglet is in. Through the aglet context object, an aglet is able to find out what other aglets are also in its current environment. Like in Telescript, a stationery aglet is used to represent the local computer's services.

## 4.4 Security

Security is a critical part of mobile agent systems. Karjoth, Lange and Oshima [KLO97] identify three security issues specific to mobile agent systems. These are:

- Protecting the host from the mobile agent,
- Protecting the mobile agent from other mobile agents, and
- Protecting the mobile agent from the host.

Researchers have so far only found solutions to the first two issues. [KLO97] [BC96]

Two major techniques are used to protect the host computer:

- Executing agents in an isolated environment. Agents cannot directly access any parts of the host system outside their execution environment. The agent system may grant some agents special privileges to access resources outside of their execution environment.
- Authenticating the source of mobile agents, and granting execution privileges to agents on the basis of how trusted their source is. Some agents may be denied execution altogether.

Java, Agent Tcl and Telescript use both of these mechanisms in their security models.

Java programs each run in their own environments.  There are security mechanisms built into the Java Virtual Machine instruction set to prevent programs from accessing outside of their environment. These are: [Ven97b]

- Type-safe reference casting.
- Structured memory access.
- Automatic garbage collection.
- Array bound checking.
- Checking references for null.

The effects of these mechanisms is that Java programs run in a sandbox. That is they are limited to the environment allocated to them by the Java Virtual Machine, and the Java byte code instruction set disallows them from directly accessing anything outside of this environment. Accesses outside of the sandbox can only be done by using some of the Java libraries, allowing disk access, network access, and printing, or by calling native methods. The Java Security Manager controls which programs are permitted access outside of the sandbox, and the nature of the outside access. For example, by default, applets are permitted to make network connections to their original source computer, but not to any other computers.   The Security Manager may grant special privileges to all classes from the same author, or to just some classes.

Agent Tcl enforces runtime security checks with a technique similar to that used by the Safe-Tcl [OLW96] interpreter.  Mobile agents are run within their own *safe* interpreters.  In the safe interpreters commands that access outside resources are hidden.  When an agent invokes a hidden command, it is redirected to the *master* interpreter.  The master interpreter implements a security policy of what commands may be available to which agents.  If the security policy allows the command for a particular agent, then the master interpreter calls the hidden command in the safe interpreter.  The security policy is user-defined by the administrator of the server.

In Telescript all agents and places have an *authority* property. The authority is a class that defines the individual or organisation in the physical world that the agent or place represents. Agents and places must reveal their authority to another agent of place on request. They may not falsify or withhold their authority. The network of places is divided into *regions* under the same authority. When an agent tries to move from one region to another, the source region must prove the authority of the agent to the destination region. [Whi96]

The Telescript language also has *permits*. Authorities limit what agents can do by assigning them permits. Permits are used to limit what instructions agents execute, and to limit their resources to a budget. For example the agent's permit can limit its lifetime or the amount of computation it may do. Telescript was designed with electronic commerce in mind, so the same resource permits can be used to allocate agents an amount of money. If an agent ever tries to violate the conditions of its permit it is destroyed. [Whi96]

The Telescript language provides a very powerful and flexible framework for protecting the host computers from untrusted sources, but at the same time not getting in the way of doing business with trusted sources.

The common way for the host to authenticate incoming mobile agents is through digital signing. Most Java mobile agent systems and Agent Tcl use this method. When an agent is transported, the message containing it is signed by the sender agent server. The receiver agent server authenticates the mobile agent message on arrival. If any part of the agent message was altered in transit, the digital signature is no longer valid. The sender agent server signs the agent rather than the original author because an agent includes the program plus the state. The state will change.

Obliq has a completely different mechanism of achieving security. Obliq relies on the lexical scoping of the semantics of the language, together with strong runtime checking. When a agent is given to a remote site for execution, because of lexical scoping these agents can only access data or resources that they can reference via free identifiers, or that are given in as procedure parameters. Lexical scoping dictates that the free identifiers refer to values that are available at the client site. Hence, the only way an agent can obtain access to a server's resources is by assigning variables to resources that the server exports to the client site. The values of these variables can then be passed as parameters to the agent. Hence, the agent is only able to access server resources that the server explicitly exports. [Car95]

The following code fragment illustrates. agent1 uses a local resource. agent2 is able to use a remote resource by obtaining a binding to an exported remote resource, and passing this as a parameter to the agent.

```
let agent1 = proc(arg)
     resource = getResource();
     use(resource)
end;
let agent2 = proc(resource, arg)
     use(resource)
end;
```
*(get a handle to remote site execution engine)*
```
let remoteSite = net_import("RemoteServer", Namer);
```

*(Execute the agent1 at the remote site -- local resource is used)*
```
remoteSite(proc (arg) agent1(arg) end)
```
*(Get resource that the remote site exports)*
```
resource = getResource(remoteSite)
```
*(Execute the agent2 at the remote site, remote resource is passed as parameter)*
```
remoteSite(proc (arg) agent2(resource, arg) end)
```

## 4.5 Efficiency

Mobile agents need to be executed reasonably efficiently. Execution performance is often not an important issue for the mobile agent itself. For agents with a high mobility rate, the bottleneck to performance is likely to be the network rather than their execution speed. Execution speed is also not critical for agents that spend most of their time idle waiting for events to happen, (such as the agent that monitors stock prices.) For such applications, even the slowest scripting languages will probably suffice. However, performance speed may be an issue for the server running the mobile agents. If the speed of the mobile agents is faster, then the server has a capacity for running more agents. Performance efficiency may also become an issue for the user. In the future, it may be that users will have to pay for the computation resources used by their mobile agents. Agents written in a more efficient language will inflict lower bills.

Java was designed to be high performance interpreted language. Java programs are compiled to Java byte code, instructions for the Java virtual machine. The byte codes are interpreted at runtime. Java programs running on Sun's implementation of the Java 1.1 virtual machine are estimated to execute at about 10 times slower than optimized native C. This is an extremely good performance for an interpreted language. [Fla97] Java's performance will be improved again with the implementation of Just-In-Time compilers. This is a technology that numerous companies are currently working on. Java byte code is compiled to native binaries just prior to program execution, giving an execution speed almost as fast as optimized native C. [jav94] The compilation however causes an overhead at the application start up. Whether Just-In-Time compilers will be useful for mobile agents depends on the application. The compilation penalty will only payoff for mobile agents that stay at one site for a relatively long time.

Tcl was not designed for performance, but as a high level scripting language for gluing components together. The runtime speed of a Tcl program is between one hundred and ten thousand times slower than optimized native C. [SBD94] However, this speed may be adequate for many mobile agent applications. There is work being done on Tcl compilers. This offers a significant speed ups to Tcl's runtime performance.  Unfortunately, the work on Tcl compilers is currently not unified with Agent Tcl.

## 4.6 Cross platform

In most cases it is desirable for a mobile agent to be able to migrate across a heterogeneous network.  Certainly, for a mobile agent to be used on the Internet this is a requirement.  For this to be possible, the agent must be written in a language that is supported on all its potential host computers.  This is one of the reasons why nearly all mobile agent systems use interpreted languages.  All the languages looked at are interpreted.

Telescript, Java and Agent Tcl agents are all interpreted at execution.  Interpreters for these languages exist across different platforms.  (Obliq interpreters are currently only available for UNIX.)  Despite this Java has a number of advantages in this area.  First, Java Virtual Machine interpreters already exist on many computers.  Most major operating system vendors, including Microsoft, Sun, IBM, Novell and Apple have announced that they plan to include the Java Virtual Machine as part of the next releases of their respective operating systems.  Mobile agents written in Java will not require a special purpose interpreter to run.  The mobile agent interpreter can be expected to be already available on most machines.  Agent Tcl requires a special purpose interpreter.  Telescript programs require a Telescript execution engine, a closed standard commercial product.  One cannot realistically expect the Telescript execution engine to become as widely spread as Java Virtual Machine interpreters.  Second, a general problem with cross platform technology is that, despite the intentions, some parts of the implementation act differently on different platforms.  While this is certainly a problem with Java now, one might optimistically expect these bugs to be fixed, simply because of the magnitude of the resources involved in Java research and development.

As a sign perhaps that General Magic accepts that Java has become the cross platform standard, it is attempting to implement a Java-based equivalent of its Telescript technology.

## 4.7 Language structure

The language that the program is written in should suit the task.  There are two views as to what is required of the task for mobile agents.  The language should be compatible with agent-oriented programming.  There is also an issue of what level of language is suitable for writing mobile agents.

Agents can be well modelled with Object oriented languages.  Agha [Agh90] argues that agents are extensions of objects.  Like objects, agents are self-contained autonomous entities.  Like objects, agents have properties and perform actions, mapping to the object-oriented concepts of attributes and methods.   The other object-oriented principles: inheritance and polymorphism are also compatible with agent programming.  Object oriented languages are well suited to representing agents.  Telescript implements agents as a built-in class.  All Telescript agents need to be derived from this class.  The various Java implementations of mobile agents also define a base agent class, from which all agents are subclasses of.

Tcl is not object oriented.  Tcl has no code modularisation other than procedures.  This is seen as a problem by the makers of Agent Tcl.  However, there is an object-oriented extension of Tcl called [incr Tcl].  The Agent Tcl developers are optimistic that they will be able to unify Agent Tcl with the object-oriented extensions. [Gra95b]

There is also an issue whether a lower level system language or a high level scripting language is more suitable to writing mobile agents.  In mobile agents languages, Java represents the system languages.  Tcl, Python and Perl represent the scripting languages.  Telescript and Obliq lie somewhere in between.  The advantage of system languages are execution speed and flexibility.  Scripting languages are well suited to gluing components together.  The advantage of scripting languages is speed of development.  For writing agents to customise the services provided on network servers, scripting languages seem to be well suited.  For lower level tasks and performance critical applications, a system language like Java is well suited.  As mobile agents become widespread it will be interesting to see which applications dominate.

Declarative languages may also be useful for writing mobile agents.  Declarative languages are well suited to knowledge representation and reasoning.  Hence they would seem suitable for writing intelligent mobile agents.  It is interesting that there have been no prominent mobile agent implementations using a declarative language.

## 5. Conclusions

Mobile agent languages are able to support the following capabilities:

- support for agent migration,
- support for agent-to-agent communication,
- support for interaction with local resources,
- security mechanisms,
- suitable execution efficiency,

- language implementation across multiple platforms, and
- ease of programming of the tasks mobile agents perform.

Of the languages considered, Telescript is arguably the best language for implementing mobile agents.  It is a language that has been designed specifically for this purpose.  The Telescript language directly addresses each of the problem specified.  The problem with Telescript is that it is proprietary software and a closed standard.

The Java language is multi-purpose, but it has necessary capabilities for writing mobile agents. Java is inferior to Telescript in the areas of support for agent migration, communication between agents and interfacing access to host computer resources.  In the other areas however Java at least equals Telescript.  Java's advantage over Telescript is that it has an open specification. What makes a mobile agent useful is the ability to run on remote machines. In the future it would seem likely that there will be many more hosts available with Java Virtual Machines than those with Telescript engines. Hence even though the Telescript language may be better than Java for writing mobile agents, Java agents will probably be able to run on more machines. The situation is in some ways analogous to Beta and VHS, (Apple Macs and PCs.)  An open standards system that delivers the same functionality to the user can be expected in the long run to gain a greater market share than a proprietary technology.

Agent Tcl is a high level scripting language that has many of Telescript's capabilities with respect to agent migration and agent communication.  Agent Tcl and Java are not in direct competition, since they offer different capabilities.

Mobile agents appear to be on the verge of entering mainstream computing.  There are currently many competing agent languages.  Only a few will gain enough support to enable the vision of mobile agents roaming the internet become a reality.